\documentclass{mn2e}
\usepackage{graphicx}

\def\ltsima{$\; \buildrel < \over \sim \;$}
\def\lsim{\lower.5ex\hbox{\ltsima}}
\def\gtsima{$\; \buildrel > \over \sim \;$}
\def\gsim{\lower.5ex\hbox{\gtsima}}

\newcommand{\be}{\begin{equation}}
\newcommand{\en}{\end{equation}}

\def\cmdue {\rm \ cm^{-2}}

\def\msole {~M_{\odot}}

\begin{document}
   \title[Probing the ambient medium of GRB 090618]{Probing the ambient medium of GRB 090618  
with XMM-Newton observations}

\author[S. Campana et al.]{S. Campana$^{1,}$\thanks{E-mail: sergio.campana@brera.inaf.it}, P. D'Avanzo$^{1}$, 
D. Lazzati$^{2}$, S. Covino$^{1,3}$, G. Tagliaferri$^{1}$, N. Panagia$^4$\\
$^1$ INAF-Osservatorio Astronomico di Brera, Via Bianchi 46, I--23807, Merate (Lc), Italy\\
$^2$ Department of Physics, NC State University, 2401 Stinson Drive, Raleigh, NC 27695-8202, USA\\
$^3$ INAF/TNG Fundaci\'on Galileo Galilei, Rambla Jos\'e Ana Fern\'andez P\'erez, 7, 38712 Bre\~na Baja, 
Tenerife, Spain\\
$^4$ Space Telescope Science Institute, 3700 San Martin Drive, Baltimore, MD21218, USA\\
}

\maketitle

\begin{abstract}
Long Gamma--ray Bursts (GRBs) signal the death of massive stars. The
afterglow emission can be used to probe the progenitor ambient through a
detailed study of the absorption pattern imprinted by the circumburst 
material as well as the host galaxy interstellar medium on the continuum spectrum. 
This has been done at optical wavelengths with impressive results.
Similar studies can in principle be carried out in the X--ray band, 
allowing us to shed light on the material metallicity, composition and distance of the absorber. 
We start exploiting this route through high resolution spectroscopy XMM-Newton observations of GRB 090618. 
We find a high metallicity absorbing medium ($Z\gsim 0.2\,Z_\odot$) with possible enhancements of S and 
Ne with respect to the other elements (improving the fit at a level of $>3.4\,\sigma$). 
Including the metallicity effects on the X--ray column density 
determination, the X--ray and optical evaluations of the absorption are in agreement for a Small 
Magellanic Cloud extinction curve.
\end{abstract}


\section{Introduction}

Tracking the evolution of metals from star-size scale to large-scale structures 
is a major step in understanding the evolution of the Universe. 
Metals are produced in stellar interiors and then ejected in their environs through 
supernova (SN) explosions and stellar winds, thus enriching the interstellar medium (ISM) of their galaxy. 
So far, the study of metal enrichment in galaxies at high redshifts has been carried out mostly 
by observing galaxies along the line of sight of bright quasars. 
This technique is plagued by selection effects, e.g. radiation from quasars probes 
more probably halos of the intervening galaxies. Gamma-ray bursts (GRBs) are opening a completely 
new window in understanding the history of metals and galaxy formation, in particular at 
high redshift. This is because, for geometrical reasons, quasar sight lines should preferentially 
intersect the outer regions of the ISM in high$-z$ galaxies. In contrast, GRBs originate within the densest 
regions of their host galaxies where massive stars are produced, probing denser environments 
(Prochaska et al. 2007).
It is now recognized that long-duration GRBs are linked to collapse of massive 
stars, based on the association between (low$-z$) GRBs and (type Ic) core-collapse supernovae 
(Woosley \& Bloom 2006 and references therein). Indeed, at least $\sim 75\%$ of GRBs with known redshift 
show intrinsic (i.e. at the redshift of the host galaxy) X--ray absorption larger than $3\times 10^{21}\cmdue$ 
(Campana et al. 2010; see also Stratta et al. 2005; Campana et al. 2006). 
In particular, more than $80\%$ of GRBs show an intrinsic column density larger than $5\times 10^{21}$
cm $^{-2}$ (Campana et al. 2010). 
Furthermore, the presence of absorption indicates a significant metal enrichment, because 
X--rays are effectively photoelectrically absorbed only by metals. Metallicities larger than 
those derived from Damped Lyman Alpha (DLA) studies at the same redshift are also measured from 
optical studies of GRB. They indicate that the metal abundance can be as high as 
$10\%$ of the solar one up to $z=6$ (Savaglio 2006; Kawai et al 2006; Campana et al. 2007).
Signatures for the presence of this material can be expected in the optical light curve of 
GRB afterglows showing up as flux enhancements in their light curves or as (variable) 
fine-structure transition lines in their spectra (Vreeswijk et al. 2007; D'Elia et al. 2009a, 2009b) 
as well as Ly$\alpha$ (Th\"one et al. 2011). 
The observations of these features can set important 
constraints on the density and distance of the absorbing material located either in the
star-forming region within which the progenitor formed or in the circumstellar environment 
of the progenitor itself (Prochaska et al. 2006; Dessauges-Zavadsky et al. 2006; 
D'Elia et al. 2009a, 2009b).

\begin{table}
\caption{XMM-Newton observation log.}
\begin{tabular}{cccc}
\hline
Instrument& Start time       & Exp. time & Net exp. time \\
                   & (hr after burst)& (s)                   & (s) \\
\hline
MOS1       & 13:48:09 (5.33)& 22615 & -- \\
MOS2       & 13:48:08 (5.33)& 22620 & -- \\
pn              &14:10:31 (5.70)& 21034 & 12695 \\
RGS1       &13:47:24 (5.32)& 22842 & 19675 \\
RGS2       &13:47:32 (5.32)& 22789 & 19651 \\
\hline
\end{tabular}
\end{table}

\begin{figure*}
\begin{center}
\includegraphics[width=13cm,angle=-90]{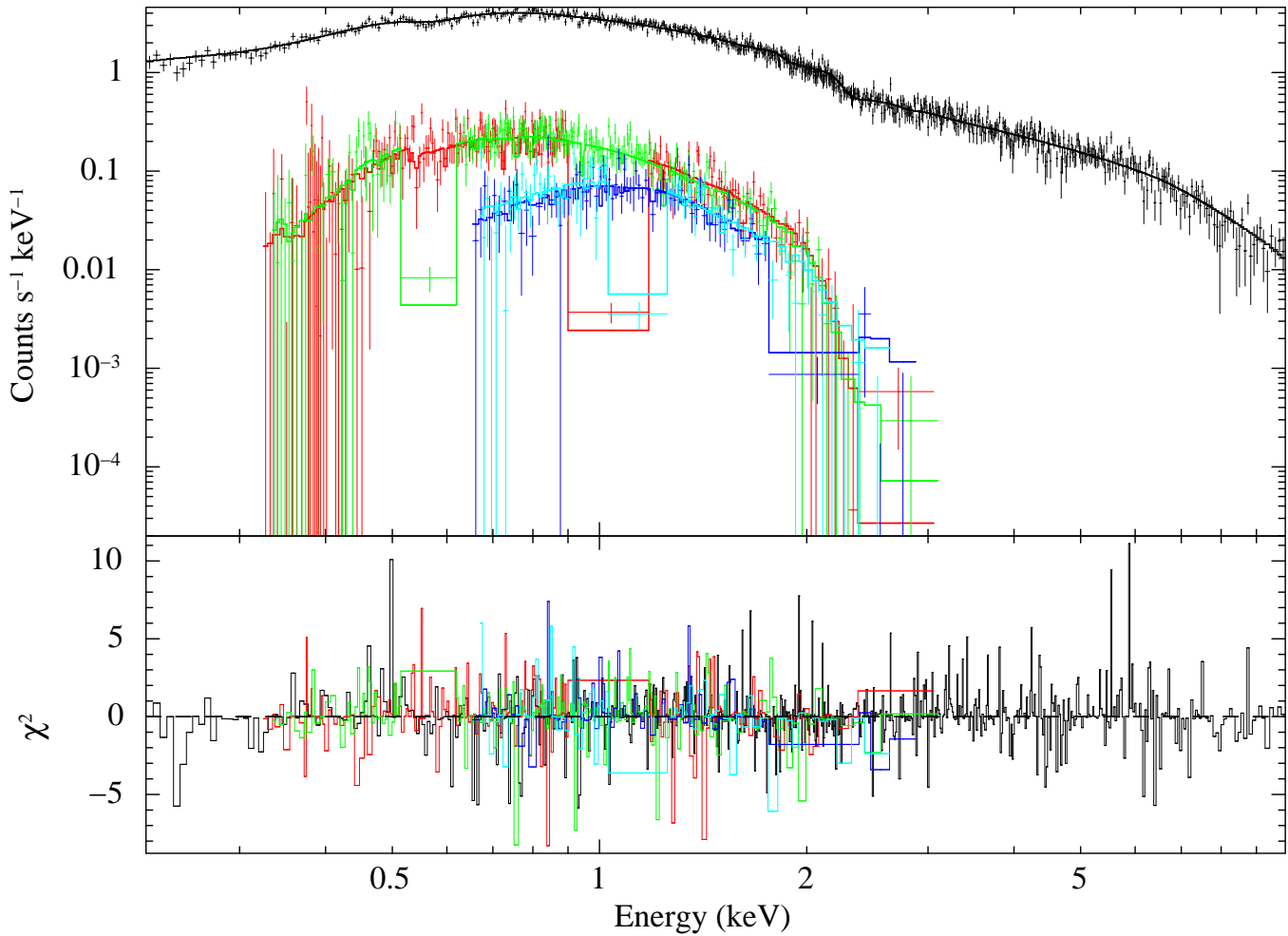}
\end{center}
\caption{XMM-Newton spectrum of GRB 090618 and X--ray continuum model.
The upper dataset (black) refers to EPIC/pn data. The middle spectra are RGS1 (red) and RGS2 (green)
first order spectra. The lower spectra are RGS1 (blue) and RGS2 (magenta) second order spectra.
The best fit power law model with intrinsic absorption with solar metallicity is shown with a solid line.
In the lower panel the contribution to the $\chi^2$ of each channel (retaining the sign) is shown.
}
\label{metal}
\end{figure*}

Metal abundance derived from optical spectroscopy can be underestimated 
because some of the metals can be locked in dust grains, the composition of which can hardly be derived through 
optical spectroscopy.  This is not the case of X--ray measurements:  X--rays are photoelectrically absorbed and 
scattered by neutral and ionised atoms as well as by dust grains, providing a full account of the metal content. 
Indeed, the soft X--ray band is the best suited for metallicity diagnostics:
absorption comes from the innermost shells (K) of elements from carbon (0.28 keV) to zinc 
(9.66 keV), with oxygen (0.54) and iron (7.11 keV) being the most prominent. 
Iron L-shell edges are also detectable. In the X--ray domain, however, relatively little 
progress has been achieved in the study of GRBs, mostly due to low statistics, 
to low spectral resolution and to the fact that some absorbing features are redshifted 
outside the instrument spectral band.
A transient feature consistent with being a Fe absorption has been reported for
GRB990705 and GRB011211 (Amati et al. 2000; Frontera et al. 2004) in the prompt emission phase. 
More recently, a decreasing total absorbing 
column density has been reported in one of the farthest GRB050904 (at $z=6.3$) and 
this has been modeled, together with optical data, to unveil the high metallicity of the 
circumburst medium (Campana et al. 2007). None of these studies however have been able to shed 
light on the chemical properties of the progenitor because the relatively poor photon statistics 
did not permit a more detailed analysis of the data.

Here we report on an XMM-Newton observation of the bright GRB 090618 started 5.3 hr after the
GRB onset. In Section 2 we describe the GRB 090618 properties. In Section 3 we describe XMM-Newton 
data extraction and analysis and in Section 4 we discuss our results.

\section{GRB 090618}

The {\it Swift} Burst Alert Telescope (BAT) detected and located GRB 090618 on June 18 2009, 08:28:29 UT 
(Schady et al. 2009). The burst was bright with a 1 second 15--150 keV peak flux of $38.9\pm0.8$ ph 
cm$^{-2}$ s$^{-1}$. 
The burst duration (as measured by the time including $90\%$ of the total flux, $T_{90}$)
in the 15--350 keV energy band is $T_{90}=113.2\pm0.6$ s. 
The GRB monitor onboard Fermi also observed this burst finding a peak energy of 
$E_p=155.5^{+11.1}_{-10.5}$ keV and a 8--1000 keV fluence of $(2.70\pm0.06) \times 10^{-4}$ erg cm$^2$ (McBreen 2009).

The X--ray telescope (XRT) started observing GRB 090618 124 s after the onset.
The burst was initially very bright also in X--rays, with a peak rate of $\sim 8,000$ 
c s$^{-1}$, and then decayed rapidly with a power-law slope of $\sim t^{-6}$. The afterglow
decay showed a break around 310 s flattening to $t^{-0.7}$. The light curve 
than breaks at $t_{X1}=5,500$ s to a steeper decay index
$t^{-1.4}$, and then again at $t_{X2}=3.1\times10^5$ s 
to a decay as $t^{-1.9}$ (Schady, Baumgartner \& Beardmore 2009).
 
The optical afterglow was detected by the P60 telescope approximately 2 minutes after the burst.
The initial magnitude was $r'\sim 13.8$ (Cenko 2009). 
The afterglow showed a decay and a rebrightening reaching a peak at 120 s with a first break 
around $\sim 600$ s (Li et al. 2009).
The afterglow was detected in all the {\it Swift}-UVOT filters (Schady 2009)
and the light curve evolution is different from what observed in the X--ray band
(Schady, Baumgartner \& Beardmore 2009).
Malesani (2009) noted that at the position of the afterglow there is a bright host galaxy with $r=22.7$ and $i=22.2$.
Cano et al. (2011) collected optical information on this burst and provided evidence for the presence of a supernova
at late times.

\begin{figure}
\begin{center}
\includegraphics[width=6.5cm,angle=-90]{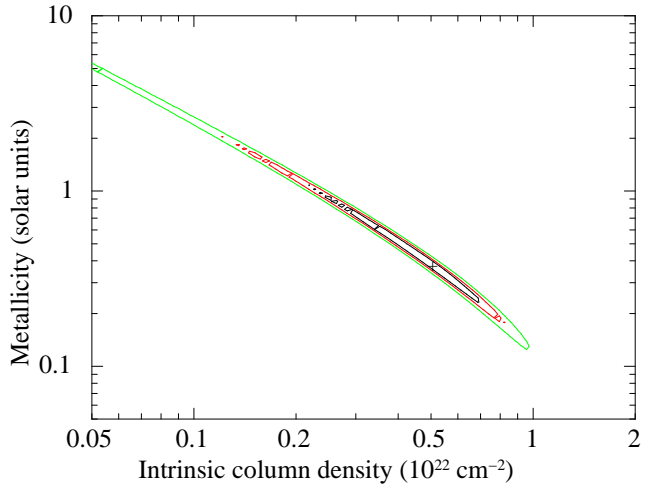}
\end{center}
\caption{Contour plot of metallicity of the absorbing medium intrinsic to the host galaxy (with fixed solar 
abundance pattern) vs. the medium column density. The cross indicates the best fit. Lines mark 
1, 2 and 3 $\sigma$ confidence levels.}
\label{metally}
\end{figure}

\begin{table*}
\caption{Spectral results.}
\begin{tabular}{ccccc}
\hline
Model        & $N_H(z)$          & Metallicity & Power law             & $\chi^2$ (dof)\\
             & ($10^{21}$)       &(solar units)& photon index          &              \\
\hline
Fix. Metall.                      &$2.3^{+0.1}_{-0.1}$&  1 (fix.)& $1.91^{+0.02}_{-0.02}$& 1288.6 (1262) \\
Free Metall.                    &$5.0^{+2.3}_{-3.0}$& $>0.2$&$1.89^{+0.02}_{-0.02}$  & 1286.8 (1261)\\
Free. Metall+Ne+S+Fe&$12^{+6}_{-8}$        & $<0.6$&$1.91^{+0.02}_{-0.02}$ & 1263.1 (1258)\\
\hline
\end{tabular}
\end{table*}

Given the afterglow brightness a spectrum has been obtained with different facilities.
The 3-m Shane telescope at Lick observatory 
observed GRB 090618 about 20 minutes after the burst. A number of strong absorption 
features identified as Mg II, Mg I, and Fe II at a common redshift of $z=0.54$ were observed 
(Cenko et al. 2009). Fatkhullin et al. (2009) confirmed the redshift based on the detection of 
an [OII] line.


At $z=0.54$ and using standard cosmology the isotropic equivalent energy in the 8--1000 keV band 
is $E_{iso} = 2.0\times 10^{53}$ erg, making GRB 090618 one of the brightest (in terms 
 of total energy emitted) burst. In addition,  at $z=0.54$
GRB 090618 is  one of the very few GRBs nearby and intrinsically very bright.

\section{XMM-Newton observations}

\subsection{Data preparation}

XMM-Newton started observing GRB 090618 on June 18, 2009 13:48:09.0 UT, 5.3 hr after the burst onset,
during $t^{-1.36}$ decay part of the afterglow light curve.
The exposure time was 22.6 ks for the MOS1 and MOS2 instruments, 21.0 ks for the pn instrument 
(all in full window mode), and 22.8 ks for the RGS1 and RGS2 instruments (see Table 1). 
Given the source strength MOS data are piled-up and were not considered in the following analysis.
Data were processed using SAS version 9.0.0, using the most 
updated calibration files available. Standard data screening criteria were applied in the extraction of 
scientific products. 
Data of the pn instrument were filtered to a 0.3--10 keV rate over the entire instrument less than 100  
c s$^{-1}$, resulting in 12.7 ks net exposure time. This filtering threshold has been selected 
as a trade-off between low background and exposure time. We did not filtered out RGS data.
Despite its strength, the GRB afterglow emission pn data were not
piled-up (less than $1\%$ at the beginning of the observation when the GRB afterglow emission was stronger). 
RGS data, thanks to the dispersion of the counts over a larger detector area, are not piled-up.
The pn data were extracted from a large circular region with radius 1,600 pixel (80 arcsec, pattern 0 to 4).
We extracted 85,000 counts for a mean count rate of 5.8 c s$^{-1}$. 
The background was extracted on the same CCD of the source from a 1,030 pixel 
radius circular region (in order to avoid the GRB streak) at $\sim 5'$ from the source. 
The background accounted for less than $10\%$ of the 
source counts. The response matrix and the ancillary response file were generated with the dedicated software tasks.

RGS data were reduced using the {\tt rgsproc} task. First and second order background subtracted spectra 
for the two instruments were retained and the appropriate response matrices were generated.
RGS data were slightly filtered in order to avoid the strongest part of the solar flares, resulting in 19.5 ks.
In total we collected 4125 (1579) for the first (second) order spectrum with RGS1 and  4381 (1504) for the first (second)
order spectrum with RGS2, respectively. The background for these spectra is around $25-40\%$ of the source flux.

\subsection{Data analysis}

The spectra are fit with a power law spectrum absorbed by a Galactic component and by an intrinsic ($z=0.54$) component.
In order to assess the contribution of the Galactic absorption we first considered HI maps which provide 
a weighted mean column density at the position of the GRB of $5.8\times 10^{20}\cmdue$ 
(Kalberla et al. 2005) and $6.5\times 10^{20}\cmdue$ (Dickey \& Lockman 1990).
To fit the absorption component we adopted the Wilms et al. (2000) model ({\tt tbabs} within the X--ray fitting
program XSPEC) and abundance pattern.
To check the Galactic absorption we fit the data of a closeby ($8'$ from the GRB)
AGN (R.A.: 19h35m57.5 Dec.: +78d21m27.0)
with a power law model. The resulting column density is $9.5^{+4.8}_{-3.8}\times 10^{20}\cmdue$ 
($90\%$ confidence level). This is in line with the predictions by HI maps even if somewhat on the high side.
Despite its small influence, we fix the Galactic contribution to $6.5\times 10^{20}\cmdue$.
We perform the first fit grouping the RGS data to 20 counts per spectral bin and the pn data to 50 counts per 
bin in order to have a good tracing of the continuum at high energies.
A fit with the metallicity fixed to the solar values gives an intrinsic column density at $z=0.54$ of 
$N_H(z)=2.3^{+0.1}_{-0.1}\times 10^{21}\cmdue$ with a $\chi^2=1288.6$ with 1262 degrees of freedom (dof, i.e. a 
reduced $\chi_{\rm red}^2=1.02$) and $29\%$ null hypothesis probability (nhp). The power law photon index is 
$\Gamma=1.91^{+0.02}_{-0.02}$ (see Table 2). We also checked that the column density did not decrease 
by splitting into two parts the entire observation with almost the same number of counts.

Given the relatively large number of counts collected we leave free to vary the absorbing medium metallicity
(keeping the solar abundance pattern). The fit do not improve significantly with $\chi^2=1286.8$ with 1261 dof 
($\chi_{\rm red}^2=1.02$) and $30\%$ nhp. 
An error search shows that the metallicity of the absorbing medium is $\gsim 0.2\,Z_\odot$ (see Fig. 2).
The intrinsic column density is $N_H(z)=5.0^{+2.3}_{-3.0}\times 10^{21}\cmdue$. For the lower limiting metallicity
the column density is $<7.7\times 10^{21}\cmdue$.

The number of counts is not high enough to afford an element by element analysis. To circumvent this we leave free 
the abundance of the single elements, one at a time, keeping the others all tied together. For many of them we just find upper 
limits (see Table 3). In particular, a few elements show an abundance not consistent with zero, namely Ne, S and Fe.

\begin{table}
\caption{Abundance limits$^*$.}
\begin{tabular}{cc|cc}
\hline
Element & Abundance & Element & Abundance\\
        & (solar units) &        & (solar units)\\
\hline
C  & $<4.9$  & N  & $<4.9$ \\
O  & $<0.6$ & Ne & $0.4-0.7$ \\
Mg & $<0.7$ & Si & $<2.2$ \\
S  & $2.6-25$ & Fe & $0.2-0.8$ \\
\hline
\end{tabular}

\noindent $^*$ Upper limits are $90\%$ confidence level for one parameter of interest ($\Delta \chi^2=2.71$).
Intervals are $90\%$ confidence level for one parameter of interest. These values were obtained by leaving free 
the abundances of each element, one by one keeping the other abundances tied together and free to vary.

\end{table}

Motivated by this analysis we fit the same spectra leaving free the abundances of Ne, S and Fe and linking together all the 
other elements (mainly driven by Oxygen). The fit improves with a $\chi^2=1263.1$ with 1258 dof 
($\chi_{\rm red}^2=1.00$) and $45\%$ nhp. Formally, an F-test gives a random probability for the 
improvement of $3.3\times 10^{-5}$, equivalent to a significance of $4.1\,\sigma$. 
The new fit provides a column density of $N_H=1.2^{+0.6}_{-0.8}\times 10^{22}\cmdue$
and just an upper limit on the metallicity of the remaining elements of $<0.6\,Z_\odot$. 
The abundances of the three elements left free to vary together lie 
within ${\rm Ne}=0.4-1.8\,{\rm Ne}_\odot$, ${\rm S}=1.5-19\,{\rm S}_\odot$,
and ${\rm Fe}<0.3\,{\rm Fe}_\odot$, respectively. A contour plot of S and Ne abundances is shown in Fig. 3.
We also searched for structures due to absorption of different ionization stages and/or to dust 
around the edges of the above elements at the known GRB redshift of $z=0.54$, finding none.
Given the presence of possible uncertainties in the Galactic column density determination, in the 
adopted solar abundance pattern we left free to vary the Galactic column density in a $\pm 50\%$ range around 
the selected value. We then repeat the fit with a constant metallicity and the fit with Ne, S and Fe free to vary.
The fit with constant metallicity is characterized by a large Galactic column density (at the edge of the allowed
distribution) and with a $\chi_{\rm red}^2=1.02$  (1260 dof). The fit with variable Ne, S and Fe abundances 
gives a $\chi_{\rm red}^2=1.00$  (1257 dof). In this case the F-test gives a a random probability for the 
improvement of $6.0\times 10^{-4}$, equivalent to a significance of $3.4\,\sigma$.

\begin{figure}
\begin{center}
\includegraphics[width=6.5cm,angle=-90]{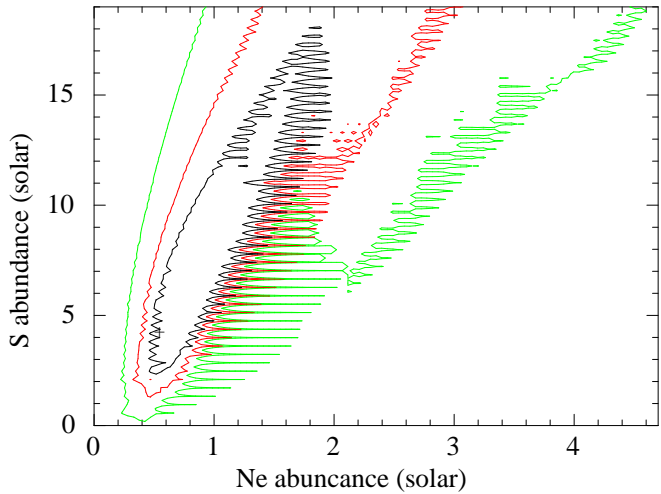}
\end{center}
\caption{Contour plot of the Ne vs. S abundances in solar units. The cross indicates the best fit. Lines mark 
1, 2 and 3 $\sigma$ confidence levels. The spiky features present in the contour plot are not real and are likely due to 
numerical problems induced by the constrained interval within which the Galactic column density lies.
The real contour level should be obtained by joining the outermost values.}
\label{nes}
\end{figure}

We also search for possible emission lines. We first search for an iron line looking for an unresolved emission line
in the 6.4--6.9 keV (rest frame) energy interval. This range is covered by pn data only. We do not find any significant 
line in this range and we are able to put an upper limit ($90\%$ confidence) of 37 eV to any line.
We also inspect the RGS data. One of the strongest (narrow) emission lines is OVIII at 654 eV (see, e.g., Bertone et al. 2010
for the list of line searched). We find a hint for the presence of this line at an energy of 432 eV (665 eV rest frame)
even if its significance is between $90\%$ and $99\%$ (see Fig. 4). The line equivalent width is 4.8 eV.
The velocity difference of the line energy with respect to the quoted redshift  is on the blue side of the spectrum, 
possibly indicating an outflow, but amounts to $5,000\pm650$ km s$^{-1}$, which seems somewhat high.

\begin{figure}
\begin{center}
\includegraphics[width=6.5cm,angle=-90]{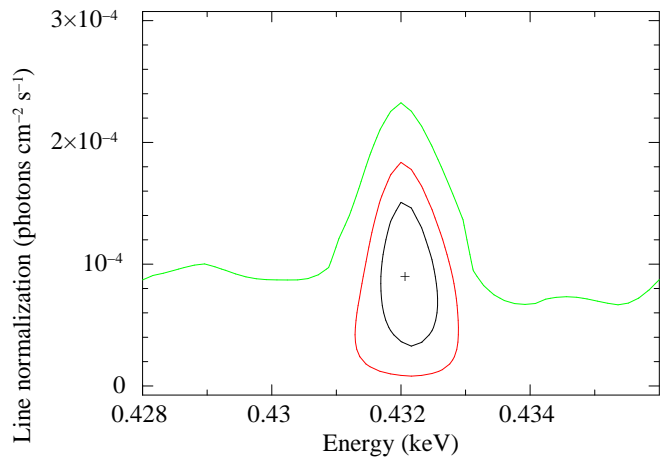}
\end{center}
\caption{Contour plot (line energy vs. line count rate normalization) of the putative O VIII detected in the XMM-Newton data.
Energies are in the observer frame. Lines mark 
1, 2 and 3 $\sigma$ confidence levels.}
\label{line}
\end{figure}

\section{Discussion and conclusions}

X--ray observations hold a great potential to probe the environment surrounding GRBs 
thanks to the imprint left by metals on the afterglow spectrum. 
Here we exploit an XMM-Newton observation of the bright GRB 090618 discovered by {\it Swift}.
The XMM-Newton afterglow spectrum is well fit by a power law model absorbed by a (known) Galactic 
component and an intrinsic component at the GRB redshift ($z=0.54$\footnote{An analysis carried out 
with free $z$ and metallicity fixed to the solar value shows that there are two solutions 
one at $z=0.5-0.7$ and another at $z\sim 3.4$. This indicates that XMM-Newton data alone, in 
this case, are not sufficient to uniquely identify the GRB redshift.}). 

Assuming the Wilms et al. (2000) solar abundance patter, we estimate the mean metallicity 
of the absorbing medium close to the GRB. 
The metallicity of the medium is strongly 
covariant with the total amount of the absorbing column (see Fig. 2) and we are able to set a 
lower limit of $\gsim 0.2\, Z_\odot$ ($90\%$ confidence level). 
The collapsar scenario for GRBs requires a massive helium star with rapidly rotating core 
(Woosley 1993; MacFadyen \& Woosley 1999). 
Low metallicities are requested by models in order to ensure that the rotationally induced 
mixing process produces a quasi-chemically homogeneous stellar evolution avoiding
the spin-down of the stellar core (Izzard, Ramirez-Riuz \& Tout 2004; 
Yoon \& Langer 2005; Woosley \& Heger 2006). This metallicity threshold is expected to 
be around $0.1-0.3\, Z_\odot$. If the medium probed by our observations has not been heavily enriched
by the progenitor and is therefore indicative of the progenitor's metallicity, our result is (barely) consistent 
with model predictions. Given however the low redshift of the GRB and the brightness of the host galaxy 
it can also be expected that the mean metallicity of the medium is high.

Analyzing the pn and RGS data we found that 
the absorbing medium might be overabundant in Ne and S (see Fig. 3 and Table 3).
Both best fit column densities are $\sim 6\times 10^{17}\cmdue$. Clearly we do not have an idea of the
distance of the absorbing medium nor of its geometry. Considering spherical symmetry and a distance scale of 0.1 pc
we have $0.01\msole$ and $0.02\msole$ mass of Ne and S, respectively, as well as $11\msole$ of hydrogen. 
This sets the order of magnitude of the distance in the case of a medium enriched by the progenitor. Larger distances 
naturally involve a pre-existing and pre-enriched medium likely by type Ib/Ic supernovae (enrichment by type II 
supernova will involve also an O overabundance).

\begin{figure}
\begin{center}
\includegraphics[width=6.5cm,angle=-90]{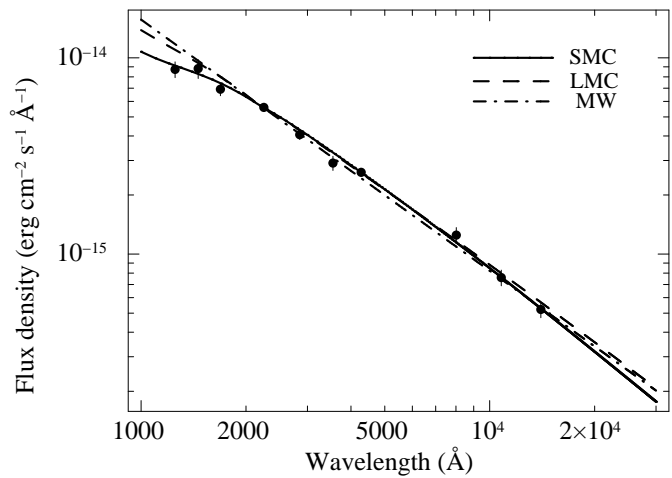}
\end{center}
\caption{Optical and UV Spectral energy distribution data fit. Data were taken and shifted according to the
overall light curve at a common time of 600 s after the burst trigger. The continuous line fit refers to an SMC 
extinction curve; the dashed line fit to the LMC and the dot-dashed line to the Milky Way.}
\label{av}
\end{figure}

An account of the absorption can be obtained by studying the spectral energy distribution (SED) from UV to nIR.
This is made possible by the detection of GRB 090618 by {\it Swift}/UVOT and by several robotic telescopes at
early times. Fitting the SED at 600 s one can infer, in addition to the Galactic absorption of $A_V=0.29$, an 
intrinsic absorption of about $A_V(z)\sim 0.13\pm0.06$, using a Small Magellanic Cloud (SMC) extinction law
($\chi^2/{\rm dof}=6.3/7$).  A fit with a Large Magellanic Cloud or Milky Way extinction laws provides 
unsatisfactory fits ($\chi^2/{\rm dof}=18.9/7$ and $\chi^2/{\rm dof}=23.2/7$).
We note that this value is somewhat smaller than the value reported by Cano et al. (2011) at later times 
of $A_V(z)\sim 0.24\pm0.09$, but consistent within the errors.

\begin{figure}
\begin{center}
\includegraphics[width=6.5cm,angle=-90]{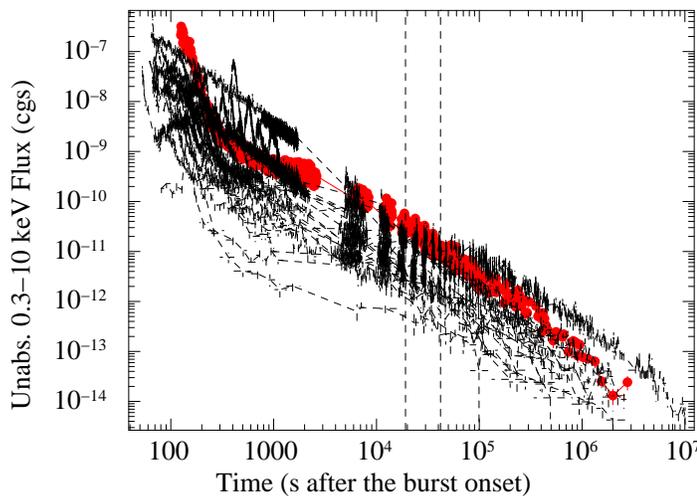}
\end{center}
\caption{Light curves of all {\it Swift} XRT GRBs with a count rate larger than 1,000 c s$^{-1}$. GRBs observed 
with XRT during the prompt event were excluded. The unabsorbed X--ray flux is in the 0.3--10 keV energy band. 
The light curve of GRB 090618 is marked with filled dots (red in the on-line version), all the other light curves with dashed lines. 
The two vertical dashed lines indicate the start and the end of the XMM-Newton observation.}
\label{line}
\end{figure}

For a SMC extinction one would expect $N_H=A_V \times 1.6 \times 10^{22}$ cm$^2$ (Weingartner \& Draine 2001). 
This implies for the solar abundance pattern fit with fixed metallicity of the X--ray spectrum a predicted optical 
absorption $A_V(z)=0.14\pm0.01$ and with free metallicity $A_V(z)<0.48$. Both estimates are
in agreement with the observations.
If we consider instead the fit with a peculiar abundance pattern (i.e. the one with Ne, S and Fe free) the 
intrinsic column density is larger and the predicted $A_V(z)=0.75^{+0.38}_{-0.50}$ is just barely consistent 
with observations (see Fig. 5). 
We note that it is not common that X--ray and optical absorption are in agreement. 

Finally, we try to put in a context our exploratory study of bright GRBs with XMM-Newton (see also Campana et 
al. 2011).
In Fig. 6 we show the light curves of {\it Swift} GRBs with a count rate larger than 1,000 c s$^{-1}$.
Excluding those GRBs observed during the prompt event also with XRT, because triggered on a precursor event,
we have 17 events. At the time of the XMM-Newton observation GRB 090618 is one of the brightest indicating that
to improve on the results obtained in this letter we would need: 
$i$) larger intrinsic column densities (in order to have larger absorption signatures). This burst has an intrinsic 
column density of $2.3\times 10^{21}\cmdue$, placing it in the tail of the column density distribution (Campana et al. 2010), 
$ii$) longer exposure times (in order to collect more photons, extending by another 25 ks the observation 
would have added about $50\%$ more photons), $iii$) faster response (in order to collect more photons, being on target 1 hr before 
would increase the total number of collected photons by $\sim 20\%$),  and $iv$) lower particle background (in order to throw away less exposure 
time and in turn get more photons). Along these lines XMM-Newton studies of the ambient medium surrounding GRBs is promising.

\section{Acknowledgments}
This work has been partially supported by ASI grant I/011/07/0.
This work made use of data supplied by the UK {\it Swift} Science Data Centre 
at the University of Leicester.

{}

\end{document}